%% file: main.tex
\documentclass[11pt,letterpaper,dvipsnames]{article}
\usepackage[T1]{fontenc}
\usepackage{gitinfo}
\usepackage{stmaryrd}
\usepackage[margin=1in]{geometry}
\usepackage{rpmacros}
\usepackage{xfrac}
\usepackage{mathpazo}
\usepackage{bm}
\usepackage{todonotes}
\usepackage{lipsum}
\usepackage{setspace}
\usepackage{scrextend}
\usepackage[shortlabels]{enumitem}
\usetikzlibrary{matrix}

\RequirePackage[pdfstartview=FitH,pdfpagemode=UseNone,colorlinks=true]{hyperref}
\hypersetup{
  linkcolor=[rgb]{0.3,0.3,0.6},
  citecolor=[rgb]{0.2, 0.6, 0.2},
  urlcolor=[rgb]{0.6, 0.2, 0.2}
}

\input{thmmacros}

\input{lazymacros} 
\input{customurlbst/bibmacros}

\newcommand{\ignore}[1]{}

\newif\ifnote
\notetrue
\ifnote
\newcommand{\RPnote}[1]{\textcolor{BrickRed}{\guillemotleft RP: #1 \guillemotright}}
\newcommand{\ATnote}[1]{\textcolor{OliveGreen}{\guillemotleft AT: #1 \guillemotright}}
\newcommand{\CRnote}[1]{\textcolor{Purple}{\guillemotleft CR: #1 \guillemotright}}
\newcommand{\MKnote}[1]{\textcolor{Orange}{\guillemotleft MK: #1 \guillemotright}}
\else
\newcommand{\RPnote}[1]{}
\newcommand{\ATnote}[1]{}
\newcommand{\CRnote}[1]{}
\newcommand{\MKnote}[1]{}
\fi

\newcommand{\ehref}[1]{\href{mailto:#1}{#1}}

\newcommand{\RSDesign}{\operatorname{RS-Design}}
\newcommand{\g}{\mathsf{Gen}}

\allowdisplaybreaks
\onehalfspace

\title{If $\VNP$ is hard, then so are equations for it}

\author{
  Mrinal Kumar\thanks{\ehref{mrinal@cse.iitb.ac.in}. Department of Computer Science \& Engineering, IIT Bombay, Mumbai, India.}
  \and
  C. Ramya\thanks{\ehref{c.ramya@tifr.res.in}. School of Technology and Computer Science, Tata Institute of Fundamental Research, Mumbai, India. Research supported by a fellowship of the DAE, Government of India.}
  \and
  Ramprasad Saptharishi\thanks{\ehref{ramprasad@tifr.res.in}. School of Technology and Computer Science, Tata Institute of Fundamental Research, Mumbai, India. Research supported by Ramanujan Fellowship of DST, and by DAE, Government of India.}
  \and
  Anamay Tengse\thanks{\ehref{anamay.tengse@gmail.com}. School of Technology and Computer Science, Tata Institute of Fundamental Research, Mumbai, India. Research supported by a fellowship of the DAE, Government of India.}%
}

\date{}

\begin{document}

\maketitle


\begin{abstract}
  
  Assuming that the Permanent polynomial requires algebraic circuits of exponential size, we show that the class $\VNP$ \emph{does not} have efficiently computable equations.
  In other words, any nonzero polynomial that vanishes on the coefficient vectors of all polynomials in the class {\sf VNP} requires algebraic circuits of super-polynomial size.

  In a recent work of Chatterjee and the authors~\cite{CKRST20}, it was shown that the subclasses of $\VP$ and $\VNP$ consisting of polynomials with bounded integer coefficients \emph{do} have equations with small algebraic circuits.
  Their work left open the possibility that these results could perhaps be extended to all of $\VP$ or $\VNP$.
  The results in this paper show that assuming the hardness of Permanent, at least for $\VNP$, allowing polynomials with large coefficients does indeed incur a significant blow up in the circuit complexity of equations.

\end{abstract}

\newpage
\section{Introduction}

In the context of proving lower bounds in complexity theory, many of the existing approaches for proving Boolean circuit lower bounds were unified by Razborov and Rudich under the {\em Natural Proofs} framework~\cite{RR97} and they showed that, under standard cryptographic assumptions, any technique that fits into this framework cannot  yield very strong lower bounds.
In the last few years there has been some work (e.g.
\cite{G15}, \cite{GKSS17, FSV18}) aimed at developing an analogue of the Natural Proofs framework for algebraic circuit lower bounds.
A crucial notion in this context is that of an \emph{equation} for a class of polynomials which we now define.

For a class ${\cal C}$ of polynomials, an {\em equation} for ${\cal C}$ is a family of nonzero polynomials such that it vanishes on the coefficient vector of polynomials in ${\cal C}$.\footnote{Strictly speaking, these notions need us to work with families of polynomials, even though we sometimes drop the word \emph{family} for ease of exposition.}
Informally, an {\em algebraic natural proof} for a class ${\cal C}$ is a family of {\em equations} for ${\cal C}$ which can be computed by algebraic circuits of size and degree polynomially bounded in their number of variables.
Thus, a lower bound for ${\cal C}$ can be proved by exhibiting an explicit polynomial on which an equation for ${\cal C}$ does not vanish.

Many of the known algebraic circuit lower bounds fit into this framework of algebraically natural proofs as observed by several authors \cite{AD08,G15,FSV18,GKSS17}, thereby motivating the question of understanding whether techniques in this framework can yield strong algebraic circuit lower bounds; in particular, whether such techniques are sufficient to separate $\VNP$ from $\VP$.
Thus, in this framework, the first step towards a lower bound for $\VP$ is to understand whether $\VP$ has a family of equations which itself is in $\VP$, that is its degree and its algebraic circuit size are polynomially bounded in the number of the variables.
The next step, of course, would be to show the existence of a polynomial family in $\VNP$ which \emph{does not} satisfy this family of equations.
This work is motivated by the first step of this framework, that is the question of understanding whether natural and seemingly rich circuit classes like $\VP$ and $\VNP$ can have efficiently constructible equations.
We briefly discuss prior work on this problem, before describing our results.

\subsection{Complexity of Equations for classes of polynomials}

In one of the first results on this problem, Forbes, Shpilka and Volk~\cite{FSV18} and Grochow, Kumar, Saks and Saraf~\cite{GKSS17} observe that the class $\VP$ does not have efficiently constructible equations if we were to believe that there are hitting set generators for algebraic circuits with sufficiently succinct descriptions.
However, unlike the results of Razborov and Rudich \cite{RR97}, the plausibility of the pseudorandomness assumption in \cite{FSV18, GKSS17} is not very well understood.
The question of understanding the complexity of equations for $\VP$, or in general any natural class of algebraic circuits, continues to remain open.

In a recent work of Chatterjee and the authors~\cite{CKRST20}, it was shown that if we focus on the subclass of $\VP$ (in fact, even $\VNP$) consisting of polynomial families with bounded integer coefficients, then we indeed have efficiently computable equations.
More formally, the main result in \cite{CKRST20} was the following.
\begin{theorem}[\cite{CKRST20}]
\label{thm:CKRST-complexes}
For every constant $c > 0$, there is a polynomial family $\{P_{N, c}\} \in \VP_\Q$ \footnote{For a field $\F$, $\VP_{\F}$ denotes the class $\VP$ where the coefficients of the polynomials are from the field $\F$.
Similarly, $\VNP_{\F}$ denotes the class $\VNP$ where the coefficients of the polynomials are from the field $\F$.}
such that for all large $n$ and $N = \binom{n+n^c}{n}$, the following are true.
  	\begin{itemize}
        \item For every family $\{f_n\} \in \VNP_\Q$, where $f_n$ is an $n$-variate polynomial of degree at most $n^c$ and coefficients in $\{-1, 0, 1\}$, we have
  		\[
    		P_{N,c}(\cvector{f_n}) = 0 \, .  
  		\] 
  		\item There exists a family $\set{h_n}$ of $n$-variate polynomials and degree at most $n^c$ with coefficients in $\{-1, 0, 1\}$ such that 
  		\[
  			P_{N,c}(\cvector{h_n}) \neq 0 \, .
    	\]
  	\end{itemize}
  	Here, $\cvector{f}$ denotes the coefficient vector of a polynomial $f$. 
\end{theorem}

Many of the natural and well studied polynomial families like the Determinant, the Permanent, Iterated Matrix Multiplication, etc., have this property of bounded coefficients, and in fact the above result even holds when the coefficients are as large as $\poly(N)$.
Thus, \autoref{thm:CKRST-complexes} could be interpreted as some evidence that perhaps we could still hope to prove lower bounds for one of these polynomial families via proofs which are algebraically natural.
Extending \autoref{thm:CKRST-complexes} to obtain efficiently constructible equations for all of $\VP$ (or even for slightly weaker models like formulas or constant depth algebraic circuits) is an extremely interesting open question.
In fact, even a conditional resolution of this problem in either direction, be it showing that the bounded coefficients condition in \autoref{thm:CKRST-complexes} can be removed, or showing that there are no such equations, would be extremely interesting and would provide much needed insight into whether or not there \emph{is} a natural-proofs-like barrier for algebraic circuit lower bounds.

\subsection{Our results}

In this paper, we show that assuming the Permanent is hard, the constraint of bounded coefficients in \autoref{thm:CKRST-complexes} is necessary for efficient equations for $\VNP$.
More formally, we show the following theorem.

\begin{restatable}[Conditional Hardness of Equations for {\sf VNP}]{theorem}{MainTheorem}
\label{thm:MainThm}
  Let $\epsilon > 0$ be a constant. Suppose, for an $m$ large enough, we have that $\Perm_{m}$ requires circuits of size $2^{m^\epsilon}$.

  Then, for $n = m^{\epsilon/4}$, any $d \leq n$ and $N = \binom{n+d}{n}$, we have that every nonzero polynomial $P(x_1,\ldots, x_N)$ that vanishes on all coefficient vectors of polynomials in $\VNP_{\C}(n,d)$  has $\operatorname{size}(P) = N^{\omega(1)}$. 
\end{restatable}
\begin{remark*}
Our proof of the above theorem easily extends to any field of characteristic zero. We shall just work with complexes for better readability.
\end{remark*}


Extending the result in \autoref{thm:MainThm} to hardness of equations for $\VP$, even under the assumption that Permanent is sufficiently hard, is an extremely interesting open question.
Such an extension would answer the main question investigated in \cite{FSV18, GKSS17} and show a natural-proofs-like barrier for a fairly general family of lower bound proof techniques in algebraic complexity.
Our proof of \autoref{thm:MainThm} however crucially relies on some of the properties of $\VNP$ and does not appear to extend to $\VP$.

Although the proof of the above theorem is quite elementary, the main message (in our opinion) is that we do not\footnote{Or rather, the results of \cite{CKRST20} and the above theorem seem to provide \emph{some} evidence for both sides!}
have compelling evidence to rule out, or accept, the efficacy of algebraic natural proofs towards proving strong lower bounds for rich classes of algebraic circuits.

\subsection{An overview of the proof}

As was observed in \cite{FSV18, GKSS17}, a lower bound for equations for a class of polynomials is equivalent to showing the existence of succinctly describable hitting sets for this class.
For our proof we show that, assuming that the permanent is sufficiently hard, the coefficient vectors of polynomials in $\VNP$ form a \emph{hitting set} for the class $\VP$.
The connection between hardness and randomness in algebraic complexity is well known via a result of Kabanets and Impagliazzo \cite{KI04}, and we use this connection, along with some additional ideas for our proof.
We briefly describe a high level sketch of our proof in a bit more detail now.

Kabanets and Impagliazzo~\cite{KI04} showed that using any explicit polynomial family $\{f_n\}$ that is sufficiently hard, one can construct a hitting set generator for $\VP$, that is, we can construct a polynomial map $\g_f:\F^{k} \rightarrow \F^{t}$ that ``fools'' any small algebraic circuit $C$ on $t$ variables in the sense that $C(y_1, y_2, \ldots, y_t)$ is nonzero if and only if the $k$-variate polynomial $C\circ\g_f$ is nonzero.
In a typical invocation of this result, the parameter $k$ is much smaller than $t$ (typically $k = \poly\log t$).
Thus, this gives a reduction from the question of polynomial identity testing for $t$-variate polynomials to polynomial identity testing for $k$-variate polynomials.
Another related way of interpreting this connection is that if $\{f_n\}$ is sufficiently hard then $\g_f$ is a polynomial map whose image does not have an equation with small circuit size.
Thus, assuming the hardness of the Permanent, this immediately gives us a polynomial map (with appropriate parameters) such that its image does not have an efficiently constructible equation.

For the proof of \autoref{thm:MainThm}, we show that the points in the image of the map $\g_{\Perm}$, can be viewed as the coefficient vectors of polynomials in $\VNP$, or, equivalently in the terminology in \cite{FSV18, GKSS17}, that the Kabanets-Impagliazzo hitting set generator is $\VNP$-succinct.
To this end, we work with a specific instantiation of the construction of the Kabanets-Impagliazzo generator where the underlying construction of combinatorial designs is based on Reed-Solomon codes.
Although this is perhaps the most well known construction of combinatorial designs, there are other (and in some parameters, better) constructions known.
However, our proof relies on the properties of this particular construction to obtain the succinct description.
Our final proof is fairly short and elementary, and is based on extremely simple algebraic ideas and making generous use of the fact that we are trying to prove a lower bound for equations for $\VNP$ and not $\VP$.

\paragraph{Details of the proof.}
Let us assume that for some constant $\epsilon > 0$ and for all\footnote{To be more precise, we should work with this condition for ``infinitely often'' $m\in \N$ and obtain that $\VNP$ does not have efficient equations infinitely often. We avoid this technicality for the sake of simplicity and the proof continues to hold for the more precise version with suitable additional care. } $m \in \N$, $\Perm_{m}$ requires circuits of size $2^{m^{\epsilon}} $.
Kabanets and Impagliazzo~\cite{KI04} showed that, for every combinatorial design $\mathcal{D}$ (a collection of subsets of a universe with small pairwise intersection) of appropriate parameters, the map
\[
  \g_{\Perm}(\vecz) = \left(\Perm(\vecz_S)\;:\; S \in \mathcal{D}\right)
\]
where $\vecz_S$ denotes the variables of in $\vecz$ restricted to the indices in $S$, is a hitting set generator for circuits of size $2^{o(m^{\epsilon})} $.
Our main goal is to construct a polynomial $F(\vecy,\vecz)$ in $\VNP$ such that
\begin{equation}
\label{eq:polyH}
F(\vecy,\vecz) = \sum\limits_{S\in {\cal D}} \operatorname{mon}_S(\vecy) \cdot {\sf Perm}(\vecz_S)
\end{equation}
By choosing parameters carefully, this would immediately imply that any equation on $N$-variables, for $N = \binom{n+d}{d}$, that vanishes on the coefficient vector of polynomials in $\VNP(n,d)$ (which are $n$-variate polynomials in $\VNP$ of degree at most $d$) requires size super-polynomial in $N$.
 
To show that the polynomial $F(\vecy,\vecz)$ in \autoref{eq:polyH} is in $\VNP$, we use a specific combinatorial design.
For the combinatorial design ${\cal D}$ obtained via Reed-Solomon codes, every set in the design can be interpreted as a univariate polynomial $g$ of appropriate degree over a finite field.
The degree of $g$ (say $\delta$) and size of the finite field (say $p$) are related to the parameters of the design ${\cal D}$.
Now,
\begin{equation}
\label{eq:polyH2}
F(\vecy,\vecz) = \sum_{\substack{g \in \F_p[v]\\ \deg(g) \leq \delta}} \inparen{\prod_{i=0}^{\delta} y_i^{g_i}} \cdot \Perm(\vecz_{S(g)}),
\end{equation}
where $(g_0,\ldots,g_\delta)$ is the coefficient vector of the univariate polynomial $g$.
Expressing $F(\vecy,\vecz)$ in \autoref{eq:polyH2} as a polynomial in $\VNP$ requires us to implement the product $\inparen{\prod\limits_{i=0}^{\delta} y_i^{g_i}}$ as a polynomial when given the binary representation of coefficients $g_0,\ldots,g_{\delta}$ via a binary vector $\vect$ of appropriate length (say $r$).
This is done via the polynomial $\Mon(\vect,\vecy)$ in \autoref{subsec:monomials} in a straightforward manner.
Furthermore, we want to algebraically implement the selection $\vecz_S$ for a set $S$ in the combinatorial design when given the polynomial $\vecg$ corresponding to $S$.
This is implemented via the polynomial $\RSDesign(\vect,\vecz)$ in \autoref{subsec:selections}.
Finally, we have
\begin{align*}
  F(\vecy,\vecz) &= \sum_{\vect \in \set{0,1}^r} \Mon(\vect, \vecy) \cdot \Perm(\RSDesign(\vect, \vecz))
\end{align*} 
which is clearly in $\VNP$ as $ \Perm_p$ is in $\VNP$ and polynomials $\Mon(\vect,\vecy)$ and $\RSDesign(\vect, \vecz)$ are efficiently computable.
We refer the reader to \autoref{section:details} for complete details.

\paragraph*{Related results.} The concept of algebraically natural
proofs was first studied in the works of Forbes, Shpilka and Volk~\cite{FSV18} and Grochow, Kumar, Saks and Saraf~\cite{GKSS17} who showed that constructing efficient equations for a class directly contradicts a corresponding {\em succinct} derandomization of the polynomial identity testing problem.
In fact, Forbes, Shpilka and Volk~\cite{FSV18} unconditionally ruled out equations for depth-three multilinear formulas computable by certain structured classes of algebraic circuits using this connection. However, this does not imply anything about complexity of equations for general classes of algebraic circuits such as $\VP$ and $\VNP$.
In the context of proving algebraic circuit lower bounds, Efremenko, Garg, Oliveira and Wigderson~\cite{EGOW18} and Garg, Makam, Oliveira and Wigderson~\cite{GMOW19} explore limitations of proving algebraic circuit lower bounds via rank based methods.
However, these results are not directly concerned with the complexity of equations for circuit classes.

Recently, Bl\"{a}ser, Ikenmeyer, Jindal and Lysikov~\cite{BIJL18} studied the complexity of equations in a slightly different context.
They studied a problem called ``matrix completion rank'', a measure for tensors that is $\NP$-hard to compute.
Assuming $\mathsf{coNP} \nsubseteq\exists\mathsf{BPP}$, they construct an explicit tensor of large (border) completion rank such that any efficient equation for the class of tensors of small completion rank must necessarily also vanish on this tensor of large completion rank.
That is, efficient equations cannot certify that this specific tensor has large (border) completion rank.
Subsequently, this result was generalized to {\em min-rank} or {\em slice-rank} ~\cite{BILPS19}.
The set-up in these papers is different from the that in our paper, and that of \cite{GKSS17,FSV18}.
One way to interpret this difference is that \cite{BIJL18} shows that ``variety of small completion rank tensors'' cannot be ``cut out'' by efficient equations, whereas the set-up of \cite{GKSS17, FSV18} and our paper would ask if \emph{every} equation for this variety requires large complexity. 

In the context of equations for varieties in algebraic complexity, Kumar and Volk~\cite{KV20} proved polynomial degree bounds on the equations of the Zariski closure of the set of non-rigid matrices as well as small linear circuits over all large enough fields.

\section{Preliminaries}

\subsection{Notation}

\begin{itemize}\itemsep0pt
\item We use $[n]$ to denote the set $\set{1,\ldots, n}$ and $\inbracket{n}$ to denote the set $\set{0,1,\ldots, n}$. We also use $\N_{\geq 0}$ to denote the set of non-negative integers.
\item We use boldface letters such as $\vecx, \vecy$ to denote tuples, typically of variables. When necessary, we adorn them with a subscript such as $\vecy_{[n]}$ to denote the length of the tuple.
\item   We also use $\vecx^{\vece}$ to denote the monomial $\prod x_i^{e_i}$.  We write $\vecx^{\leq d}$ for the set of all monomials of degree at most $d$ in $\vecx$, and $\F[\vecx]^{\leq d} $ for the set of polynomials in $\vecx$ over the field $\F$ of degree at most $d$.
\item As usual, we identify the elements of $\F_p$ with $\{0, 1, \ldots, p-1\}$ and think of $\inbracket{n}$ as a subset of $\F_p$ in the natural way for any $n < p$.
\end{itemize}



\subsection{Some basic definitions}

\subsubsection*{Circuit classes}

\begin{definition}[Algebraic circuits] \label{defn:alg-circuits}
  An \emph{algebraic circuit} is specified by a directed acyclic graph, with leaves (indegree zero; also called \emph{inputs}) labelled by field constants or variables, and internal nodes labelled by $+$ or $\times$.
  The nodes with outdegree zero are called the \emph{outputs} of the circuit.
  Computation proceeds in the natural way, where inductively each $+$ gate computes the sum of its children and each $\times$ gate computes the product of its children.

  The \emph{size} of the circuit is defined as the number of nodes in the underlying graph.
\end{definition}

\begin{definition}[$\VP$ and $\VNP$]
  A family of polynomials $\set{f_n}$, where $f_n$ is $n$-variate, is said to be in $\VP$ if $\deg(f_n)$ and the algebraic circuit complexity of $f_n$ is bounded by a polynomial function of $n$. That is, there is a constant $c \geq 0$ such that for all large enough $n$ we have $\deg(f_n), \operatorname{size}(f_n) \leq n^c$.

  A family of polynomials $\set{f_n}$ is said to be in $\VNP$ if there is a family $\set{g_n(\vecx_{[n]}, \vecy_{[m]})} \in \VP$ such that $m$ is bounded by a polynomial function of $n$ and 
  \[
    f_n(\vecx) = \sum_{\vecy \in \set{0,1}^m} g_n(\vecx, \vecy).\qedhere
  \]
\end{definition}

For some $n, d \in \N$, let $\mathcal{C}_{n,d}$ be a class of $n$-variate polynomials of \emph{total} degree at most $d$.
That is, $\mathcal{C}_{n,d} \subseteq \F[\vecx]^{\leq d}$.
Similarly, we will use $\VP(n,d)$ and $\VNP(n,d)$ to denote the intersection of $\VP$ and $\VNP$ respectively, with $\F[\vecx_{[n]}]^{\leq d}$.

\subsubsection*{Equations and succinct hitting sets}

\begin{definition}[Equations for a class]
\label{def:proofs}
For $N = \binom{n + d}{n}$, a nonzero polynomial $P_N(\vecZ)$ is called an {\em equation} for $\mathcal{C}_{n,d}$ if for all $f(\vecx) \in \mathcal{C}_{n,d}$, we have that $P_{N}(\cvector{f}) = 0$, where $\cvector{f}$ is the coefficient vector of $f$.
\end{definition}

Alternatively, we also say that a polynomial $P(\vecZ)$ {\em vanishes} on the coefficient vectors of polynomials in class $\mathcal{C}$ if $P_{N}(\cvector{f}) = 0$ for all $f\in \mathcal{C}$.

\begin{definition}[Hitting Set Generator (HSG)]
\label{def:HSG}
A polynomial map $G:\mathbb{F^\ell}\rightarrow \mathbb{F}^n$ given by $G(z_1,\ldots,z_\ell)=(g_1(\vecz),\ldots,g_n(\vecz))$ is said to be a {\em hitting set generator (HSG)} for a class $\mathcal{C}\subseteq \mathbb{F}[\vecx]$ of polynomials if for all nonzero $P\in \mathcal{C}$, $P\circ G = P(g_1,\ldots,g_n)\not \equiv 0$.
\end{definition}

We review the definition of {\em succinct hitting sets} introduced \cite{GKSS17, FSV18}.

\begin{definition}[Succinct Hitting Sets for a class of polynomials~\cite{GKSS17,FSV18}]
\label{def:succinct-HS}
 For $N = \binom{n + d}{n}$, we say that a class of $N$-variate polynomials \emph{$\mathcal{D}_N$ has $\mathcal{C}_{n,d}$-succinct hitting sets} if for all nonzero $P(\vecZ) \in \mathcal{D}_{N}$, there exists some $f \in \mathcal{C}_{n,d}$ such that $P_{N}(\cvector{f}) \neq 0$.
\end{definition}

\subsubsection*{Hardness to randomness connection}

For our proofs, we will need the following notion of combinatorial designs, which is a collection of subsets of a universe with small pairwise intersection.

\begin{definition}[Combinatorial designs]
A family of sets $\{S_1,\ldots,S_N\} \subseteq [\ell]$ is said to be an $(\ell,m,n)$-design if
\begin{itemize}
\item $|S_i|=m$ for each $i \in [n]$
\item $|S_i \cap S_j| < n$ for any $i \neq j$. \qedhere
\end{itemize}
\end{definition}

Kabanets and Impagliazzo \cite{KI04} obtain hitting set generators from polynomials that are hard to compute for algebraic circuits.
The following lemma is crucial to the proof of our main theorem.

\begin{lemmawp}[HSG from Hardness \cite{KI04}]
\label{lem:KI-HSG-from-hardness}
Let $\{S_1,\ldots,S_N\}$ be an $(\ell,m,n)$-design and $f(\vecx_m)$ be an $m$-variate, individual degree $d$ polynomial that requires circuits of size $s$.
Then for fresh variables $\vecy_{\ell}$, the polynomial map $\operatorname{KI-gen}_{(N,\ell,m,n)}(f) : \F^{\ell} \rightarrow \F^n $ given by
\begin{equation}\label{eqn:KI}
  \inparen{f(\vecy_{S_1}), \ldots, f(\vecy_{S_N})}
\end{equation}
is a hitting set generator for all circuits of size at most $\inparen{\frac{s^{0.1}}{N(d+1)^n}} $.
\end{lemmawp}

\section{Proof of the main theorem}

\subsection*{Notation}
\label{section:details}
\begin{enumerate}
  \item For a vector $\vect = (t_1,\ldots, t_r)$, we will use the short-hand $t_{i,j}^{(a)}$ to denote the variable $t_{(i \cdot a + j + 1)}$.
  This would be convenient when we consider the coordinates of $\vect$ as blocks of length $a$.

  \item For integers $a,p$, we shall use $\Mod(a,p)$ to denote the unique integer $a_p \in [0,p-1]$ such that $a_p = a\bmod{p}$. 
\end{enumerate}

As mentioned in the overview, the strategy is to convert the hitting set generator given in \eqref{eqn:KI} into a succinct hitting set generator. Therefore, we would like to associate the coordinates of \eqref{eqn:KI} into coefficients of a suitable polynomial. That is, we would like to build a  polynomial in $\VNP$ of the form
\[
g(y_1,\ldots, y_\ell, z_1,\ldots, z_t) = \sum_{m \in \vecy^{\leq d}} m \cdot f(\vecz_{S_m})
\]
with the monomials $m \in \vecy^{\leq d}$ suitably indexing into the sets of the combinatorial design. The above expression already resembles a $\VNP$-definition and with a little care this can be made effective. We will first show that the different components of the above expression can be made succinct using the following constructions. 

\subsection{Building monomials from exponent vectors}
\label{subsec:monomials}

For $n,r \in \N$, let $a = \floor{r/n}$, and define $\Mon_{r,n}(\vect,\vecy) $ as follows. 
\[
  \Mon_{r,n}(t_1,\ldots, t_r, y_1, \ldots, y_n) = \prod_{i=0}^{n-1} \prod_{j=0}^{a-1} \inparen{t_{i,j}^{(a)} y_{i+1}^{2^{j}} + (1 - t_{i,j}^{(a)}) }
\]
The following observation is now immediate from the definition above.
\begin{observation}
  For any $(e_1,\ldots, e_n) \in \inbracket{d}^n$, we have
  \[
    \Mon_{r,n}(\Bin(e_1),\ldots, \Bin(e_n), y_1, \ldots, y_n) = y_1^{e_1} \cdots y_n^{e_n},
  \]
  where $\Bin(e)$ is the tuple corresponding to the binary representation of $e$, and $r = n \cdot \ceil{\log_2 d}$.
  Furthermore, the polynomial $\Mon_{r,n}$ is computable by an algebraic circuit of size $\poly(n,r)$.
\end{observation}

\subsection{Indexing Combinatorial Designs Algebraically}
\label{subsec:selections}
\newcommand{\Sel}{\operatorname{Sel}}

Next, we need to effectively compute the hard polynomial $f$ on sets of variables in a combinatorial design, indexed by the respective monomials.
We will need to simulate some computations modulo a fixed prime $p$. The following claim will be helpful for that purpose.

\begin{claim}
\label{claim:interpolation}
For any $i,b,p \in \N_{\geq 0}$ with $i\leq p$, there exists a unique univariate polynomial $Q_{i,b,p}(v)\in \mathbb{Q}[v]$ of degree at most $b$ such that
\[
  Q_{i,b,p}(a) = \begin{cases}
    1 & \text{if $0 \leq a < b$ and $a \equiv i~(\bmod{p})$},\\
    0 & \text{if $0 \leq a < b$ and $a \not\equiv i~(\bmod{p})$}.
  \end{cases}
\]
\end{claim}

\begin{proof}
  We can define a unique univariate polynomial $Q_{i,b,p}(v)$ satisfying the conditions of the claim via interpolation to make a unique univariate polynomial take a value of $0$ or $1$ according to the conditions of the claim.
  Since, there are $b$ conditions, there always exists such a polynomial of degree at most $b$.
\end{proof}

For any $n,b,p \in \N_{\geq 0}$ with $n\geq p$, define
\[
  \Sel_{n,b,p}(u_1,\ldots, u_n, v) \triangleq \sum_{i=1}^n u_i\cdot  Q_{i,b,p}(v).
\]

\begin{observation}
  For any $n,b,p\in \N_{\geq 0}$ with $n\geq p$, for any $0 \leq a < b$, we have that
  \[
    \Sel_{n,b,p}(u_1,\ldots, u_n, a) = u_{\Mod(a,p)} = u_{a\bmod{p}}
  \]
  The degree of $\Sel_{n,b,p}$ is at most $(b+1)$ and can be computed by an algebraic circuit of size $\poly(b)$. 
\end{observation}

\begin{proof}
  From the definition of the univariate polynomial $Q_{i,b,p}(v)$ of degree $b$ in \autoref{claim:interpolation}, $Q_{i,b,p}(a)$ outputs $1$ if and only if $i= a \bmod{p}$.
Hence, $\Sel_{n,b,p}(u_1,\ldots, u_n, a)$ is $u_{a\bmod{p}}$ and is of degree at most $(b+1)$.
\end{proof}

\medskip
\noindent
And finally, we choose a specific combinatorial design to instantiate \autoref{lem:KI-HSG-from-hardness} with. 

\subsection{Reed-Solomon based combinatorial designs}\label{sec:RS-designs}

For any prime $p$ and any choice of  $a \leq p$, the following is an explicit construction of a $(p^2, p, a)$-combinatorial design of size $p^a$,  defined as follows:
\begin{quote}
  With the universe $U = \F_p \times \F_p$,
  for every univariate polynomial $g(t) \in \F_p[t]$ of degree less than $a$, we add the set $S_g = \setdef{(i,g(i))}{i\in \F_p}$ to the collection. 
\end{quote}
Since any two distinct univariate polynomials of degree less than $a$ can agree on at most $a$ points, it follows that the above is indeed a $(p^2, p, a)$-design.\\

The advantage of this specific construction is that it can be made succinct as follows.
For $r = a \cdot \floor{\log_2 p}$, let ${t_1,\ldots,t_r}$ be variables taking values in $\{0,1\}$.
The values assigned to $\vect$-variables can be interpreted as a univariate over $\F_p$ of degree $< a$ by considering $\vect \in \{0,1\}^r$ as a matrix with $a$ rows and $\floor{\log_2p}$ columns each~\footnote{Working with $\floor{\log_2 p}$ bits (as opposed to $\ceil{\log_2 p}$) makes the proofs much simpler, and does not affect the size of the design by much.}.
The binary vector in each row represents an element in $\mathbb{F}_p$. We illustrate this with an example.
\begin{center}
  \begin{tikzpicture}[transform shape]
    
    \node (label_t) at (-3.25,0) {$\vect = $};
    \matrix [matrix of math nodes, left delimiter=(, right delimiter=)](g) at (-1.75,0) {
      1 & 1 & 1\\
      0 & 1 & 0\\
      0 & 0 & 1\\
      1 & 0 & 0\\
      0 & 1 & 1\\
    };
    \node (convert) at (0.5,0) {$\longrightarrow$};
    \matrix [matrix of math nodes, left delimiter=(, right delimiter=)] (eg) at (2.25,0) {
      7\\
      2\\
      1\\
      4\\
      2\\
    };
    \node (label_g) at (3.75,0) {$\cong g(v)$};
    \node (info1) at (0.25,-2) {For $p = 11$, $a = 5$, $g(v) = 7 + 2v + v^2 + 4v^3 + 2v^4 \in \F_{11}[v]$,};
    \node (info2) at (0.25,-2.75) {$\vect$ is a $5 \times 3$ matrix that encodes the coefficients of $g(v)$.};

\end{tikzpicture}
\end{center}

Let $\vecz$ denote the $p^2$ variables $\set{z_{1},\ldots,z_{p^2}}$, put in into a $p \times p$ matrix.
Let $S$ be a set in the Reed-Solomon based $(p^2,p,a)$-combinatorial design.
We want to implement the selection $\vecz_S$ algebraically.
In the following, we design a vector of polynomials that outputs the vector of variables $\inparen{z_{0,g(0)\bmod{p}}^{(p)}, \ldots, z_{p-1,g(p-1)\bmod{p}}^{(p)} }$.
Note that as mentioned above the polynomial $g$ can be specified via variables $t_1,\ldots,t_r$.
That is,
\begin{align*}
  \RSDesign_{p,a}(t_1,\ldots, t_r, z_{1},\ldots, z_{p^2}) & \in (\F[\vect, \vecz])^p \quad,\quad \text{for $r = a \cdot \floor{\log_2 p}$},\\
  \RSDesign_{p,a}(t_1,\ldots, t_r, z_{1},\ldots, z_{p^2})_{i+1} &= \Sel_{p, p^3, p}\inparen{z_{i,0}^{(p)},\ldots, z_{i,p-1}^{(p)}, R_{i,a,p}(\vect)},\quad\text{for each $i \in \F_p $,}\\
  \text{where }R_{i,a,p}(\vect) & = \sum_{j=0}^{a-1} \insquare{\inparen{\sum_{k=0}^{\ell_p - 1} t_{j,k}^{(\ell_p)} \cdot 2^{k}} \cdot \Mod(i^{j}, p)},\\
  \text{with }\ell_p & = \floor{\log_2 p}.
\end{align*}

\begin{observation}
  For any prime $p$, $a\leq p$, and $\vect \in \set{0,1}^r$ for $r = a \cdot \floor{\log_2p}$, we have
  \[
    \RSDesign_{p,a}(\vect, \vecz) = \inparen{z_{i,g(i)}\;:\; i \in \F_p},
  \] where $g(v) \in \F_p[v]$ is the univariate whose coefficient vector is represented by the bit-vector $\vect$.
  Furthermore, the polynomial $\RSDesign_{p,a}$ is computable by an algebraic circuit of size $\poly(p)$.
\end{observation}
\begin{proof}
  Fix some $\vect \in \set{0,1}^r$.
From the definition of $R_{i,a,p}(\vect)$, it is clear that $R_{i,a,p}(\vect)$ returns an integer $\alpha$ such that $g(i) = \alpha\bmod p$ where $\vect$ encodes the coefficients of the polynomial $g(t)$ in binary.
Furthermore, since $\Mod(i^j,p)$ is the unique integer $c \in [0,p-1]$ with $c = i^j\bmod{p}$, it also follows that $R_{i,a,p}(\vect)$ is an integer in the range $[0,p^3]$.
Hence,
  \[
    \Sel_{p,p^3,p}\inparen{z_{i,0}^{(p)},\ldots, z_{i,p-1}^{(p)}, R_{i,a,p}(\vect)} = z_{i,g(i)}
  \]
  as claimed. 
\end{proof}

\subsection{The \texorpdfstring{$\VNP$}{\sf VNP}-Succinct-KI generator}

We are now ready to show the $\VNP$-succinctness of the Kabanets-Impagliazzo hitting set generator when using a hard polynomial from $\VNP$ and a Reed-Solomon based combinatorial design.

For a prime $p$ and for the largest number $m$ such that $m^2 \leq p $, we will use $\Perm_{[p]} \in \F[\vecy_{[p]}]$ to denote $\Perm_m$ applied to the first $m^2 $ variables of $\vecy$.

We now define the polynomial $F_{n,a,p}(\vecy_{[n]},\vecz_{[p^2]}) $ as follows.
\begin{align}
\label{eqn:VNP-poly}
  F_{n,a,p}(y_1,\ldots, y_n,z_1,\ldots, z_{p^2}) &= \sum_{\vect \in \set{0,1}^r} \Mon_{r,n}(\vect, \vecy) \cdot \Perm_{[p]}(\RSDesign_{p,a}(\vect, \vecz))\\
  \text{where }r & = a \cdot \floor{\log_2 p} \nonumber
\end{align}
It is evident from the above definition that the polynomial $F_{n,a,p}(\vecy,\vecz)$ is in $\VNP$ for any $p$ that is $\poly(n)$, when seen as a polynomial in $\vecy$-variables with coefficients from $\C[\vecz]$.

From the construction, we have that
\[
  F_{n,a,p}(y_1,\ldots, y_n, z_1,\ldots z_{p^2}) = \sum_{\vece} \vecy^{\vece} \cdot \Perm_{[p]}(\vecz_{S_\vece}),
\]
where $\set{S_\vece}$ is an appropriate ordering of the Reed-Solomon based $(p^2, p,a)$-combinatorial design of size $p^a$, described in \autoref{sec:RS-designs}. 

\subsection{Putting it all together}
We are now ready to show that if the Permanent polynomial is exponentially hard, then any polynomial $P$ that vanishes on the coefficient vectors of all polynomials in the class {\sf VNP} requires super-polynomial size to compute it.

\MainTheorem*

\begin{proof}
  Let $p$ be the smallest prime larger than $m^2$; we know that $p \leq 2m^2$. We will again use $\Perm_{[p]} \in \F[\vecy_{[p]}]$ to denote $\Perm_m$ acting on the first $m^2 $ variables of $\vecy$. Therefore, if $\Perm_m$ requires size $2^{m^\epsilon}$ then so does $\Perm_{[p]}$.\\
  Consider the polynomial $F_{n,n,p}(\vecy_{[n]}, \vecz_{[p^2]}) \in \VNP$ defined in \eqref{eqn:VNP-poly}, which we interpret as a polynomial in $\vecy$ with coefficients in $\C[\vecz]$.
  The individual degree in $\vecy$ is at least $d$, and at most $p$.\\
  Let $F_{n,n,p}^{\leq d}(\vecy_{[n]},\vecz_{[p^2]})$ denote the polynomial obtained from $F_{n,n,p}$ by discarding all terms whose total degree in $\vecy$ exceeds $d$.
  By standard homogenisation arguments, it follows that $F^{\leq d}_{n,n,p} \in \VNP$ as well.
  Therefore,
  \[
    F^{\leq d}_{n,n,p}(\vecy, \vecz) = \sum_{\deg(\vecy^\vece) \leq d} \vecy^{\vece} \cdot \Perm_{[p]}(\vecz_{S_\vece}),
  \]
  where $S_{\vece}$, for various $\vece$, is an appropriate indexing into a $(p^2, p, n)$-combinatorial design of size $N$.
  Since the individual degree in $\vecy$ of $F_{n,n,p}$ was at least $d$, every coefficient of $F^{\leq d}_{n,n,p}$ is $\Perm_{[p]}(\vecz_{S})$ for some $S$ in the combinatorial design.
  In other words, the coefficient vector of $F^{\leq d}_{n,n,p}$ is precisely $\operatorname{KI-gen}_{N,p^2, p, n}(\Perm_{[p]})$.

  Suppose $P(x_1,\dots, x_N)$ is a nonzero equation for $\VNP(n,d)$, then in particular it should be zero on the coefficient vector of $F^{\leq d}_{n,n,p}(\vecy,\veca) \in \VNP$ for any $\veca \in \C^{p^2}$. By the Polynomial Identity Lemma~\cite{O22,DL78,Z79,S80}, this implies that $P$ must be zero on the coefficient vector of $F^{\leq d}_{n,n,p}(\vecy,\vecz) \in (\C[\vecz])[\vecy]$, where coefficients are formal polynomials in $\C[\vecz]$. 
  Since the coefficient vector of $F^{\leq d}_{n,n,p}(\vecy, \vecz)$ is just $\operatorname{KI-gen}_{N,p^2, p, n}(\Perm_{[p]})$, the contrapositive of \autoref{lem:KI-HSG-from-hardness} gives that
  \begin{align*}
    \size(P) & > \frac{\size(\Perm_{[p]})^{0.1}}{N \cdot 2^n} > \frac{\size(\Perm_{m})^{0.1}}{N \cdot 2^n}\\
    \implies \size(P) & > \frac{2^{0.1 m^{\epsilon}}}{N \cdot 2^n}
  \end{align*}
  Since $N = \binom{n+d}{n} \leq 2^{2n} \leq 2^{o(m^\epsilon)}$, it follows that $\size(P) = N^{\omega(1)}$.
 
 \end{proof}


  \paragraph*{Concluding that {\sf VNP} has no efficient equations}
  Note that for a family $\set{P_N}$ to be a \emph{family of equations} for a class $\mathcal{C}$, we want that for \emph{all large enough} $n$, the corresponding polynomial $P_N$ should vanish on the coefficient vectors of all $n$-variate polynomials in $\mathcal{C}$.
  This condition is particularly important if we want to use equations for $\mathcal{C}$ to prove lower bounds against it, since a family of polynomials $\set{f_n}$ is said to be computable in size $s(n)$ if $\operatorname{size}(f_n) \leq s(n)$ for \emph{all large enough} $n$.
  
  \autoref{thm:MainThm} shows that, for $m$ large enough,  if there is a constant $\epsilon > 0$ such that $\size(\Perm_m) \geq 2^{m^{\epsilon}}$, then for $n = m^{\epsilon/4}$ and any $d\leq n$, the coefficient vectors of polynomials in $\VNP(n,d)$ form a hitting set for all $N$-variate polynomials (where $N= \binom{n+d}{d}$) of degree $\poly(N)$   that are computable by circuits of size $\poly(N)$.
  Now suppose the Permanent family is $2^{m^{\epsilon}}$-hard for a constant $\epsilon > 0$, which means that $\Perm_m$ is $2^{m^{\epsilon}}$-hard for \emph{infinitely many} $m \in \N$.
  Then using \autoref{thm:MainThm}, we can conclude that for any family $\set{P_N} \in \VP$, we must have for \emph{infinitely many} $n$ that $P_N(\cvector{f_n}) \neq 0$ for some $f_n \in \VNP$, which then shows that $\set{P_N}$ is not a family of equations for $\VNP$.

\section{Discussion and Open Problems}

In the context of proving circuit lower bounds, and in relation to the notion of \emph{algebraically natural proofs}, an interesting question that emerges from the recent work of Chatterjee and the authors~\cite{CKRST20} (stated in \autoref{thm:CKRST-complexes}) is whether the condition of ``small coefficients'' is necessary for efficiently constructible equations to exist, especially for the class $\VP$.
While this question remains open for $\VP$, our result shows that this additional restriction on the coefficients is essentially vital for the existence of efficiently constructible equations for the class $\VNP$, and therefore provides strong evidence \emph{against} the existence of efficient equations for $\VNP$.

In light of \autoref{thm:CKRST-complexes} and \autoref{thm:MainThm} for $\VNP$, one could make a case that equations for $\VP$ might also incur a super-polynomial blow up, without the restriction on coefficients.
On the other hand, it could also be argued that an analogue of \autoref{thm:MainThm} may not be true for $\VP$, since our proof crucially uses the fact that $\VNP$ is ``closed under exponential sums''.
In fact, our proof essentially algebraises the intuition that coefficient vectors of polynomials in $\VNP$ ``look random'' to a polynomial in $\VP$, \emph{provided that} $\VNP$ was exponentially more powerful than $\VP$.

Thus, along with the previously known results on efficient equations for polynomials in  $\VP$ with bounded coefficients, our result highlights that  the existence of such equations for $\VP$ in general continues to remain an intriguing mystery.

\subsection*{Open Problems}

We now conclude with some possible directions for extending our results.

\begin{itemize}
\item Perhaps the most interesting question here is to prove an analogue of \autoref{thm:MainThm} for equations for $\VP$.
This would provide concrete evidence for the possibility that we cannot hope to prove very strong lower bounds for algebraic circuits using proofs which proceed via efficiently constructible equations, from a fairly standard complexity theoretic assumption.

\item At the moment, we cannot rule out the possibility of there being efficient equations for $\VP$ in general; it may be possible that the bounded coefficients condition in \autoref{thm:CKRST-complexes} can be removed. In particular, the question of proving upper bounds on the complexity of equations for $\VP$ is also extremely interesting, even if one proves such upper bounds under some reasonable complexity theoretic assumptions.
A first step perhaps would be to prove upper bounds on the complexity of potentially simpler models, like formulas, algebraic branching programs or constant depth circuits. From the works of Forbes, Shpilka and Volk~\cite{FSV18}, we know that such equations for structured subclasses of $\VP$ (like depth-$3$ multilinear circuits) cannot be \emph{too} simple (such as sparse polynomials, depth-$3$ powering circuits, etc.). Can we prove a non-trivial upper bound for equations for these structured classes within $\VP$? 
  
\item Another question of interest would be to understand if the hardness assumption in \autoref{thm:MainThm} can be weakened further.
For instance, is it true that $\VNP$ does not have efficiently constructible equations if $\VP \neq \VNP$, or if $\Perm_n$ requires circuits of size $n^{\poly\log(n)}$?
The current proof seems to need an exponential lower bound for the Permanent.
\end{itemize}

\subsection*{Acknowledgements}
We thank an anonymous reviewer of FOCS 2020, and Joshua Grochow, whose questions pointed us in the direction of this result.
We also thank Prerona Chatterjee and Ben Lee Volk for helpful discussions at various stages of this work.

\bibliographystyle{customurlbst/alphaurlpp}
\bibliography{masterbib/references,masterbib/crossref}

\end{document}

%% file: thmmacros.tex
\usepackage{amsthm}
\usepackage{thmtools,thm-restate}

\numberwithin{equation}{section}
\declaretheoremstyle[bodyfont=\it,qed=\qedsymbol]{noproofstyle}

\declaretheorem[numberlike=equation]{observation}

\declaretheorem[name=Observation,numbered=no]{observation*}

\declaretheorem[numberlike=equation]{theorem}

\declaretheorem[name=Theorem,numbered=no]{theorem*}

\declaretheorem[name=Lemma,numbered=no]{lemma*}
\declaretheorem[numberlike=equation,style=noproofstyle,name=Lemma]{lemmawp}

\declaretheorem[name=Corollary,numbered=no]{corollary*}

\declaretheorem[name=Proposition,numbered=no]{proposition*}

\declaretheorem[numberlike=equation]{claim}
\declaretheorem[name=Claim,numbered=no]{claim*}

\declaretheorem[name=Conjecture,numbered=no]{conjecture*}

\declaretheorem[name=Question,numbered=no]{question*}

\declaretheoremstyle[bodyfont=\it,qed=$\lozenge$]{defstyle} 

\declaretheorem[numberlike=equation,style=defstyle]{definition}
\declaretheorem[unnumbered,name=Definition,style=defstyle]{definition*}

\declaretheorem[unnumbered,name=Example,style=defstyle]{example*}

\declaretheorem[unnumbered,name=Notation=defstyle]{notation*}

\declaretheorem[unnumbered,name=Construction,style=defstyle]{construction*}

\declaretheorem[unnumbered,name=Remark,style=defstyle]{remark*}


%% file: lazymacros.tex
\renewcommand{\phi}{\varphi}
\renewcommand{\epsilon}{\varepsilon}

\newcommand{\size}{\operatorname{size}}

\newcommand{\vecZ}{\mathbf{Z}}

\newcommand{\Perm}{\operatorname{Perm}}

\newcommand{\cvector}[1]{\overline{\operatorname{coeff}}(#1)}

\newcommand{\C}{\mathbb{C}}

\newcommand{\inbracket}[1]{\llbracket #1 \rrbracket}

\newcommand{\Mon}{\operatorname{Mon}}
\newcommand{\Bin}{\operatorname{Bin}}
\newcommand{\Mod}{\operatorname{Mod}}


%% file: customurlbst/bibmacros.tex
\usepackage{nth}
\usepackage{intcalc}
\usepackage{etoolbox}
\usepackage{xstring}

\usepackage{ifpdf}
\ifpdf
\else
\usepackage[quadpoints=false]{hypdvips}
\fi

\newcommand{\ECCCurl}[2]{http://eccc.hpi-web.de/report/\ifnumcomp{#1}{>}{93}{19}{20}#1/#2/}

\newcommand{\shortECCC}[2]{\texttt{\href{http://eccc.hpi-web.de/report/\ifnumcomp{#1}{>}{93}{19}{20}#1/#2/}{eccc:TR#1-#2}}}

\newcommand{\parseECCC}[1]{
\StrSubstitute{#1}{TR}{}[\tmpstring]%
\IfSubStr{\tmpstring}{/}{ 
\StrBefore{\tmpstring}{/}[\ecccyear]%
\StrBehind{\tmpstring}{/}[\ecccreport]%
}{
\StrBefore{\tmpstring}{-}[\ecccyear]%
\StrBehind{\tmpstring}{-}[\ecccreport]%
}%
\shortECCC{\ecccyear}{\ecccreport}}

%% file: main.bbl
\newcommand{\etalchar}[1]{$^{#1}$}
\begin{thebibliography}{GMOW19}

\bibitem[AD08]{AD08}
Scott Aaranson and Andrew Drucker.
\newblock \href {https://www.scottaaronson.com/blog/?p=336} {Arithmetic natural
  proofs theory is sought}.
\newblock Shtetl Optimized: Scott Aaranson's Blog, 2008.

\bibitem[BIJL18]{BIJL18}
Markus Bl{\"{a}}ser, Christian Ikenmeyer, Gorav Jindal, and Vladimir Lysikov.
\newblock \href {http://dx.doi.org/10.1145/3188745.3188832} {Generalized matrix
  completion and algebraic natural proofs}.
\newblock In {\em Proceedings of the 50th Annual {ACM} {SIGACT} Symposium on
  Theory of Computing, {STOC} 2018, Los Angeles, CA, USA, June 25-29, 2018},
  pages 1193--1206. {ACM}, 2018.

\bibitem[BIL{\etalchar{+}}19]{BILPS19}
Markus Bl{\"{a}}ser, Christian Ikenmeyer, Vladimir Lysikov, Anurag Pandey, and
  Frank{-}Olaf Schreyer.
\newblock \href {http://arxiv.org/abs/1911.02534} {Variety Membership Testing,
  Algebraic Natural Proofs, and Geometric Complexity Theory}.
\newblock {\em CoRR}, abs/1911.02534, 2019.

\bibitem[CKR{\etalchar{+}}20]{CKRST20}
Prerona Chatterjee, Mrinal Kumar, C.~Ramya, Ramprasad Saptharishi, and Anamay
  Tengse.
\newblock \href {https://arxiv.org/abs/2004.14147} {On the Existence of
  Algebraically Natural Proofs}.
\newblock {\em CoRR}, abs/2004.14147, 2020.
\newblock Pre-print available at \href {http://arxiv.org/abs/2004.14147}
  {\path{arXiv:2004.14147}}.

\bibitem[DL78]{DL78}
Richard~A. DeMillo and Richard~J. Lipton.
\newblock \href {http://dx.doi.org/10.1016/0020-0190(78)90067-4} {{A
  Probabilistic Remark on Algebraic Program Testing}}.
\newblock {\em Information Processing Letters}, 7(4):193--195, 1978.

\bibitem[EGOW18]{EGOW18}
Klim Efremenko, Ankit Garg, Rafael Oliveira, and Avi Wigderson.
\newblock \href {http://dx.doi.org/10.4230/LIPIcs.ITCS.2018.1} {Barriers for
  Rank Methods in Arithmetic Complexity}.
\newblock In {\em 9th Innovations in Theoretical Computer Science Conference,
  {ITCS} 2018, January 11-14, 2018, Cambridge, MA, {USA}}, volume~94 of {\em
  LIPIcs}, pages 1:1--1:19. Schloss Dagstuhl - Leibniz-Zentrum f{\"{u}}r
  Informatik, 2018.

\bibitem[FSV18]{FSV18}
Michael~A. Forbes, Amir Shpilka, and Ben~Lee Volk.
\newblock \href {http://dx.doi.org/10.4086/toc.2018.v014a018} {Succinct Hitting
  Sets and Barriers to Proving Lower Bounds for Algebraic Circuits}.
\newblock {\em Theory of Computing}, 14(1):1--45, 2018.

\bibitem[GKSS17]{GKSS17}
Joshua~A. Grochow, Mrinal Kumar, Michael~E. Saks, and Shubhangi Saraf.
\newblock \href {http://arxiv.org/abs/1701.01717} {Towards an algebraic natural
  proofs barrier via polynomial identity testing}.
\newblock {\em CoRR}, abs/1701.01717, 2017.
\newblock Pre-print available at \href {http://arxiv.org/abs/1701.01717}
  {\path{arXiv:1701.01717}}.

\bibitem[GMOW19]{GMOW19}
Ankit Garg, Visu Makam, Rafael Oliveira, and Avi Wigderson.
\newblock \href {http://dx.doi.org/10.1109/FOCS.2019.00054} {More Barriers for
  Rank Methods, via a "numeric to Symbolic" Transfer}.
\newblock In {\em 60th {IEEE} Annual Symposium on Foundations of Computer
  Science, {FOCS} 2019, Baltimore, Maryland, USA, November 9-12, 2019}, pages
  824--844. {IEEE} Computer Society, 2019.

\bibitem[Gro15]{G15}
Joshua~A. Grochow.
\newblock \href {http://dx.doi.org/10.1007/s00037-015-0103-x} {Unifying Known
  Lower Bounds via Geometric Complexity Theory}.
\newblock {\em Comput. Complex.}, 24(2):393--475, 2015.

\bibitem[KI04]{KI04}
Valentine Kabanets and Russell Impagliazzo.
\newblock \href {http://dx.doi.org/10.1007/s00037-004-0182-6} {{D}erandomizing
  Polynomial Identity Tests Means Proving Circuit Lower Bounds}.
\newblock {\em Computational Complexity}, 13(1-2):1--46, 2004.
\newblock \pSTOC{2003}.

\bibitem[KV20]{KV20}
Mrinal Kumar and Ben~Lee Volk.
\newblock \href {https://arxiv.org/abs/2003.12938} {A Polynomial Degree Bound
  on Defining Equations of Non-rigid Matrices and Small Linear Circuits}.
\newblock {\em CoRR}, abs/2003.12938, 2020.
\newblock Pre-print available at \href {http://arxiv.org/abs/2003.12938}
  {\path{arXiv:2003.12938}}.

\bibitem[Ore22]{O22}
{\O}ystein Ore.
\newblock {\"{U}}ber h{\"{o}}here Kongruenzen.
\newblock {\em Norsk Mat. Forenings Skrifter}, 1(7):15, 1922.

\bibitem[RR97]{RR97}
Alexander~A. Razborov and Steven Rudich.
\newblock \href {http://dx.doi.org/10.1006/jcss.1997.1494} {Natural Proofs}.
\newblock {\em J. Comput. Syst. Sci.}, 55(1):24--35, 1997.

\bibitem[Sch80]{S80}
Jacob~T. Schwartz.
\newblock \href {http://dx.doi.org/10.1145/322217.322225} {{F}ast
  {P}robabilistic {A}lgorithms for {V}erification of {P}olynomial
  {I}dentities}.
\newblock {\em Journal of the ACM}, 27(4):701--717, 1980.

\bibitem[Zip79]{Z79}
Richard Zippel.
\newblock \href {http://dx.doi.org/10.1007/3-540-09519-5_73} {Probabilistic
  algorithms for sparse polynomials}.
\newblock In {\em Symbolic and Algebraic Computation, {EUROSAM} '79, An
  International Symposiumon Symbolic and Algebraic Computation}, volume~72 of
  {\em Lecture Notes in Computer Science}, pages 216--226. Springer, 1979.

\end{thebibliography}
